\documentclass[aps,prl,twocolumn,showpacs,superscriptaddress]{revtex4}
\usepackage{graphicx}
\usepackage{mathrsfs}
\usepackage{amsmath,amssymb}
\begin{document}

\title{SU(4) Kondo Effect in Carbon Nanotubes}
\author{Manh-Soo Choi}%
\affiliation{Department of Physics, Korea University, Seoul 136-701, Korea}%
\author{Rosa L\'opez}%
\affiliation{D\'epartement de Physique Th\'eorique, Universit\'e de
  Gen\`eve, CH-1211 Gen\`eve 4, Switzerland}%
\author{Ram\'on Aguado}%
\affiliation{Teor\'{\i}a de la Materia Condensada, Instituto de Ciencia
  de Materiales de Madrid (CSIC) Cantoblanco,28049 Madrid, Spain}%
\date{\today}%
\begin{abstract}
We investigate theoretically the non-equilibrium transport
properties of carbon nanotube quantum dots.  Owing to the
two-dimensional band structure of graphene, a double orbital
degeneracy plays the role of a pseudo-spin, which is entangled
with the spin.  Quantum fluctuations between these four degrees of
freedom result in an SU(4) Kondo effect at low temperatures. This
exotic Kondo effect manifests as a four-peak splitting in the
non-linear conductance when an axial magnetic field is applied.
\end{abstract}

% 73.23.-b    Electronic transport in mesoscopic systems
% 72.15.Qm Scattering mechanisms and Kondo effect
% 73.63.Fg   Electronic transport in nanotubes
% 75.20.Hr   Local moment in compounds and alloys;
% Kondo effect, valence fluctuations, heavy fermions
\maketitle

%%%%%%
\let\up=\uparrow%
\let\down=\downarrow%
\let\eps=\epsilon%
\let\veps=\varepsilon%
\newcommand\orbit{\mathrm{orb}}%
\newcommand\total{\mathrm{tot}}%
\newcommand\nm{\,\mathrm{nm}}%
\newcommand\K{\,\mathrm{K}}%
\newcommand\eV{\,\mathrm{eV}}%
\newcommand\mK{\,\mathrm{mK}}%
\newcommand\meV{\,\mathrm{meV}}%
\newcommand\half{\frac{1}{2}}%
\newcommand\bfS{\mathbf{S}}%
\newcommand\bfT{\mathbf{T}}%
\newcommand\bfsigma{{\boldsymbol{\sigma}}}%
\newcommand\bftau{{\boldsymbol{\tau}}}%
\newcommand\varH{\,\mathscr{H}}%
\newcommand\varN{\mathscr{N}}%

%%%%%%
\emph{Introduction}.--- Carbon nanotubes (CNTs) exhibit a good deal of
remarkable transport phenomena including quantum
interference~\cite{Lian01}, Luttinger liquid features~\cite{Bock99} or
spin polarized transport~\cite{Tsu99}.  Finite-length CNTs behave like
quantum dots (QDs) and thus show Coulomb blockade~\cite{qdcnt,4shellCB}
and Kondo physics~\cite{cntkondo}.  Interestingly, the richness of the
band structure of CNTs and the feasibility to attach new materials as
electrodes, e.g., ferromagnetic~\cite{Tsu99} or superconducting
contacts~\cite{contacts}, allows us to explore new aspects of the Kondo
effect, one of the most fundamental topics in condensed matter physics.

The electronic states of a CNT form one-dimensional electron and
hole sub-bands. They originate from the quantization of the
electron wavenumber perpendicular to the nanotube axis,
$k_{\perp}$, which arises when graphene is wrapped into a cylinder
to create a CNT. By symmetry, for a given sub-band at
$k_{\perp}=k_0$ there is a second degenerate sub-band at
$k_{\perp}=-k_0$. Semiclassically, this degeneracy corresponds to
the clockwise ($\circlearrowright$) or counterclockwise
($\circlearrowleft$) symmetry of the wrapping modes.
Linear transport measurements %of CNT QDs
in the Coulomb blockade
regime reveal a distinct four-fold shell filling pattern owing to
the orbital and the spin degeneracies~\cite{4shellCB}.

In this Letter we combine several theoretical approaches, scaling
theory, numerical renormalization group (NRG)~\cite{Wilson},
non-crossing approximation (NCA)~\cite{nca}, equation-of-motion
(EOM)~\cite{eom-nca} methods, to present a unified picture of
low-temperature, non-equilibrium transport through CNT quantum
dots in the presence of magnetic fields. We show that quantum
fluctuations between the four states $\{ \circlearrowright
\uparrow ,\circlearrowright \downarrow , \circlearrowleft\uparrow
, \circlearrowleft\downarrow \}$ may dominate transport at low
temperatures provided that both the orbital and spin indexes are
conserved during tunneling. This leads to a highly symmetric SU(4)
Kondo effect, and hence an enhanced Kondo temperature, in which
the spin and the orbital degrees of freedom are totally entangled
\cite{otherSU4}. We also point out that the orbital degeneracy in
the dot itself is not enough for having SU(4) Kondo physics. In
general, %multi-level
SU(2) Kondo physics is possible.  We show
that neither an enhanced Kondo temperature nor linear conductance
measurements can distinguish between the two effects. Instead, the
non-linear conductance in the presence of a \emph{parallel}
magnetic field shows a four-peak structure, with different
splittings for the spin and the orbital sectors, which
unambiguously signals SU(4) Kondo physics. Our theoretical results
are in good agreement with recent experiments by Jarillo-Herrero
\textit{et al}~\cite{Jarillo-Herrero04z}.

\emph{Model}.--- We study a quantum dot CNT coupled to left ($L$) and
right ($R$) electrodes. We consider the case where the QD has two
(nearly) degenerate localized orbitals [labeled by the quantum number
$m=1,2$ where $1(2)$ denotes $\circlearrowright(\circlearrowleft)$
orbital mode, respectively].
\begin{figure}
\centering%
\includegraphics*[width=80mm]{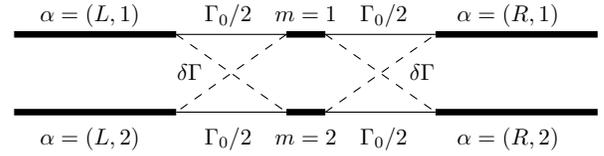}
\caption{A schematic of couplings between the degenerate orbitals
  [$m=1(2)$] with conduction channels [$L(R),m$].}%
%% Here, $\delta\Gamma=\pi\rho_0|V_{X}|^2$ denotes the cross coupling
%% constant and $\Gamma_0=\pi\rho_0|V_{0}|^2$ is the coupling constant for
%% orbital-conserving processes.
\label{su4cnt::fig:1}
\end{figure}
The presence of an axial magnetic field ($B_{\parallel}$) lifts {\it
  both} the orbital and spin degeneracies: A parallel magnetic field
induces an Aharonov-Bohm phase $2\pi\Phi/\Phi_0$, where $\Phi=\pi
d_{t}^2/4B_{\parallel}$ is the flux threading the nanotube,
$\Phi_0=h/e$ is the flux quantum and $d_{t}$ the tube diameter.
This Aharonov-Bohm flux shifts the allowed $k_\perp$ and the
orbital degeneracy is lifted by an amount $\pm\Delta_{\orbit}=\pm
ed_{t}v_F B_{\parallel}/4$, where $v_F$ if the Fermi
velocity~\cite{Ando}. The states near the energy gap correspond to
semiclassical orbits which have an \emph{orbital} magnetic moment
$\mu_{\orbit}=ed_{t}v_F/4$. Thus, an axial magnetic field leads to
an energy shift $\Delta_{\rm orb}=\mu_{\rm orb} B_{\parallel}$.
The states further split due to the Zeeman energy
$\pm\Delta_{Z}/2=\pm g\mu_B B_{||}/2$, where $\mu_B$ is the Bohr
magneton and $g\approx 2$ is the $g$-factor in nanotubes.
$\mu_\orbit$ scales with the CNT diameter and is typically one
order of magnitude larger than the Bohr
magneton~\cite{Jarillo-Herrero04z,Minot04a} resulting in a
stronger influence of $B_{\parallel}$ on the orbital sector. Thus,
the single particle energy $\epsilon_{m\sigma}$ associated with
the orbital $m$ and the spin $\sigma$ is given by: $\eps_{m\sigma}
= \eps_d + \Delta_{\orbit}(\delta_{m,1}-\delta_{m,2}) +
(\Delta_Z/2)(\delta_{\sigma,\up}-\delta_{\sigma,\down})$. The QD
is then described by the Hamiltonian
\begin{equation}
\label{TwoLevelApp::eq:HD}
\varH_D =
\sum_{m=1,2}\sum_{\sigma=\up,\down}\eps_{m,\sigma} \,
d_{m,\sigma}^\dag d_{m,\sigma}
+ \half U(n-n_g)^2 \,,
\end{equation}
where
\begin{math}
n = \sum_{m,\sigma}d_{m,\sigma}^\dag d_{m,\sigma}
\end{math}
is the occupation and $U$ is the Hubbard-like on-site interaction
(we focus on the regime where the QD is occupied by a single
electron). The two leads are modelled as:
\begin{equation}
\label{TwoLevelApp::eq:HC}
\varH_C = \sum_{\alpha\in
L,R}\sum_{m=1,2}\sum_{k,\sigma}\eps_{\alpha,k}\,
c_{\alpha,k,m,\sigma}^\dag c_{\alpha,k,m,\sigma} \,,
\end{equation}
Without loss of generality, we assume that there are two
distinguished (groups of) channels $m=1,2$ in each lead. The
coupling between the leads and the dot is described by the
tunneling Hamiltonian of the following form
\begin{equation}
\label{TwoLevelApp::eq:HT} \varH_T =
\sum_{\alpha,k,\sigma}\sum_{m,m'} V_{m,m'}^\alpha
\left(c_{\alpha,k,m,\sigma}^\dag d_{m',\sigma} + h.c.\right) \,,
\end{equation}
%For simplicity, we have ignored the $k$- and $\sigma$-dependence
%of the tunneling amplitude $V_{m,m'}^\alpha$ in
%Eq.~(\ref{TwoLevelApp::eq:HT}).
where the tunneling amplitudes read
\begin{math}
V_{m,m'}^\alpha= [V_0\delta_{m,m'} +
V_X(1-\delta_{m,m'})]/\sqrt{2},
\end{math}
(for simplicity, we ignore the $k$- and $\sigma$-dependence of the
tunneling amplitudes). In the most general case, they describe two
types of tunneling processes (see Fig.~\ref{su4cnt::fig:1}): (i)
those in which the orbital quantum number is conserved, denoted by
$V_{1,1}^\alpha=V_{2,2}^\alpha=V_0/\sqrt{2}$ and (ii) those events
accounting for mixing (cross coupling)
$V_{m,m'}^\alpha=V_X/\sqrt{2}$ with $m\neq m'$. %%
To gain more physical intuition when $V_X\neq 0$ we rewrite the
total Hamiltonian $\varH=\varH_{D}+\varH_{C}+\varH_{T}$ in terms
of symmetric (even) and antisymmetric (odd) combinations of the
%localized and delocalized
orbital channels. Before, we simplify the algebra by performing a
canonical transformation, $c_{L(R),k,m,\sigma}= (a_{k,m,\sigma}\pm
b_{k,m,\sigma})/\sqrt{2}$, such that the resulting Hamiltonian
contains only a single lead,
$a_{k,m,\sigma}=(c_{L,k,m,\sigma}+c_{R,k,m,\sigma})/\sqrt{2}$,
with two channels $m=1,2$. Next, we apply the even-odd
transformation $a_{k,1(2),\sigma}=(c_{k_{e}\sigma}\pm i
c_{k_{o}\sigma})/\sqrt{2}$ and $d_{1(2)\sigma}=(d_{e\sigma}\pm i
d_{o\sigma})/\sqrt{2}$, such that $\varH$ reads,
\begin{multline}
\label{even-odd}
\varH=\sum_{\sigma,\nu=e,o}\epsilon_{k_{\nu}}c_{k_\nu,\sigma}^\dag
c_{k_\nu,\sigma}+ \sum_{\sigma,\nu=e,o}\eps_{\nu\sigma}
d_{\nu,\sigma}^\dag d_{\nu,\sigma} + U (n-n_g)\\
+V^e\sum_{k_e,\sigma}\left( c_{k_e,\sigma}^\dag d_{e\sigma}+h.c \right)
+ V^o\sum_{k_o,\sigma} \left( c_{k_o,\sigma}^\dag d_{o\sigma} +
  h.c.\right)\,,
\end{multline}
with $V^e\equiv V_{0}+V_{X}$ and $V^o\equiv V_{0}-V_{X}$ (note
that $\epsilon_{k_\nu}$ and $\epsilon_{\nu\sigma}$ remain
invariant under this transformation). In the absence of cross
coupling terms $V_{X}=0$ both, the even and odd orbitals are
equally coupled to the QD ($V^e=V^o=V_0$). Thus, at small energies
the effective model leads to SU(4) Kondo physics, see below.
However, for the maximal mixing, i.e, $V_{X}=V_{0}$, the even
orbital is doubly coupled to the dot ($V^e=2V_{0}$) whereas the
odd orbital becomes {\it uncoupled} ($V^o=0$). Here, SU(2) Kondo
physics arises owing to spin fluctuations in the even orbital
channel.

\emph{Effective Kondo model}.---Let us now substantiate our previous
arguments by examining the low-energy properties of the system. After
performing a Schrieffer-Wolf transformation to $\varH$ the effective
Hamiltonian reads:
\begin{multline}
\label{kondo-su4su2}
\varH_{K} = \varH_{C}+
\frac{J_1}{4}\left[\bfS\cdot(\psi^\dag\bfsigma\psi) +
  \bfS\cdot(\psi^\dag\bfsigma\tau^z\psi) T^x\right]\\
+ \frac{J_2}{4} \left[\bfS\cdot(\psi^\dag\bfsigma\bftau^\bot\psi)
  \cdot\bfT^\bot +
  (\psi^\dag\bftau^\bot\psi)\cdot\bfT^\bot\right]\\
+ \frac{J_3}{4} (\psi^\dag \tau^z\psi)T^x + \frac{J_4}{4} \left[
  \bfS\cdot(\psi^\dag\bfsigma\tau^z\psi) +
  \bfS\cdot(\psi^\dag\bfsigma\psi) T^x
\right]\\
- J_5 T^x + \mu_B g \bf{B} \cdot \bf{S}+\mu_{\orbit} \bf{B} \cdot \mathbf{T},
\end{multline}
where
\begin{math}
\psi_k^\dag =[c_{e,k,\up}^\dag\; c_{e,k,\down}^\dag\;
c_{o,k,\up}^\dag\; c_{o,k,\down}^\dag]
\end{math}
are the field operators for the leads in the even-odd basis and
$\psi=\sum_k \psi_k$.  The field operator for the QD
\begin{math}
\psi_d^\dag = [d_{e,\up}^\dag\; d_{e,\down}^\dag\;
d_{o,\up}^\dag\; d_{o,\down}^\dag]
\end{math}
defines the orbital pseudo-spin and the spin operators given by
$\mathbf{T}=\psi_d^\dag{\bftau}\psi_d$ and
$\mathbf{S}=\psi_d^\dag{\bfsigma}\psi_d$ respectively, with $\bf{\tau}$
and $\bf{\sigma}$ being the Pauli matrices in the orbital pseudo-spin
and spin spaces, respectively. Here, $T^\bot(\tau^\bot)$ denotes
$T^{y}+T^{z}(\tau^{y}+\tau^{z})$.
Using renormalization group (RG) arguments, the effective coupling
constants in the $U$-large limit are given initially by $J_1 =
J_3= \varN V^2/\epsilon_d$, $J_2 = J_1 (|V_0|^2-|V_X|^2)/V^2$,
$J_4 = J_1 |2V_0 V_X|/V^2$, where $V^2=|V_0|^2+|V_X|^2$($\varN$ is
the degeneracy). $J_{5}$ is not renormalized in the RG procedure
and does not flow into the strong coupling regime. At zero
magnetic field with only spin and orbital conserving tunneling
processes ($V_{X}=0$), $J_1=J_2=J_3=J$ while $J_4=J_5=0$.  The
corresponding Hamiltonian is reduced to the SU($\varN$=4) Kondo
model where the spin $\bf{S}$ and the orbital pseudo-spin $\bf{T}$
are entangled (last term of the equation),
\begin{multline}
\label{kondo-su4} \varH_{\rm SU(4)} = \varH_C +
({J}/{4})[\mathbf{S}\cdot(\psi^\dag{\bfsigma}\psi) +
(\psi^\dag{\bftau}\psi)\cdot\mathbf{T}\\
+\mathbf{S}\cdot(\psi^\dag{\bfsigma}{\bftau}\psi)\cdot\mathbf{T}].
\end{multline}
The scaling equations are reduced to a single equation
\begin{equation}
\label{scaling1}
dJ/\ln{D}=\varN\rho_0J^2,
\end{equation}
where $\rho_0$ is the density of states (DOS) in the leads and $D$
is the bandwidth, resulting in the exponentially enhanced Kondo
temperature
\begin{math}
T_K^\mathrm{SU(4)} \approx D\exp\left[1/(\varN\rho_0J)\right]
\end{math}
%(notice the factor $\varN$ in the exponent)
compared with $T_K^\mathrm{SU(2)}$ of the single-level
SU($\varN$=2) model.
% (SLSU(2)).

In the other limiting case ($V_X=V_{0}$), the corresponding
Kondo-like Hamiltonian [Eq.~(\ref{kondo-su4su2}) at $B=0$ with
$J_1=J_3=J_4=2\varN|V_{0}|^2/\epsilon_d$ and $J_2=0)$] involves
only spin fluctuations in the doubly degenerate ($\varN$=2) even
orbital.  It gives rise to what we call a two-level (TL) SU(2)
model:
\begin{math}
\varH_\mathrm{TLSU(2)} = \varH_C +
J\bfS_{e}\cdot(\psi_{e}^\dag\bfsigma\psi_{e})\left(1+T^x\right) + (J/4)
(\psi_{e}^\dag\psi_{e})T^x - J_5 T^x
\end{math}.
The RG equation for $J$ is
\begin{equation}
\label{scaling2}
dJ/\ln D = 2\varN\rho_0 J^2.
\end{equation}
This model corresponds to an SU(2) Fermi liquid~\cite{Zarand95}
with a Kondo temperature $T_K^\mathrm{TLSU(2)}$ which is
{\emph{the same} as $T_K^\mathrm{SU(4)}$ (compare
Eqs.~(\ref{scaling1}) and (\ref{scaling2}), the factor 2 comes
from the doubling of the coupling $2V_{0}$~\cite{Boese02}).
%; see also Eq.~(\ref{even-odd})].

The scaling arguments above are confirmed by the NRG studies of the
spectral density $A_{m\sigma}(\omega)$ for the localized level
$m\sigma$.
At $B_{\parallel}=0$, the spectral density
%\emph{per spin and orbital}
shows a peak near the Fermi energy, corresponding to the
formation of the SU(4) Kondo state; see Fig.~\ref{su4cnt::fig:2}
(solid line in the right inset). The peak width, which is much
broader than that for the SU(2) Kondo model (dotted line),
demonstrates the exponential enhancement of the Kondo temperature
mentioned above. Another remarkable effect is that the SU(4) Kondo
peak shifts away from $\omega=E_F=0$ and is pinned at
$\omega\approx T_K^\mathrm{SU(4)}$.  This can be understood from
the Friedel sum rule~\cite{Langreth66a} which, in this case, gives
$\delta=\pi/4$ for the scattering phase shift at $E_F$.
Accordingly, the linear conductance at zero temperature is given
by $\mathcal{G}_0 = 4(e^2/h)\sin^2\delta = 2e^2/h$. It is
interesting to recall that the Friedel sum rule gives the same
linear conductance also for the TL SU(2) Kondo model. Thus,
neither the enhancement of the Kondo temperature nor the linear
conductance, can distinguish between the SU(4) and the TL SU(2)
Kondo effects.  This can only be achieved by studying the
influence of a parallel magnetic field, which we do now.

\emph{SU(4) Kondo model at finite field}.---
%The effects of an axial
%magnetic field $B_\|$ on the SU(4) Kondo model are stressed by the
%four-peak splitting of the Kondo peak in the spectral density
%$A_\total(\omega)$; see Fig.~\ref{su4cnt::fig:2}.  This feature has been
%observed experimentally~\cite{Jarillo-Herrero04z} and is remarkably
%simple to understand.
Because of the underlying SU(4) symmetry, the orbital pseudo-spin
should behave the same way as the real spin.  In particular, the
lift of the pseudo-spin degeneracy will split the Kondo peak (as
long as the lift is larger than the Kondo temperature) just like
the Zeeman splitting of the real spin does.  The only difference
is that pseudo-spin is more susceptible to the magnetic field than
the real spin since $\mu_\orbit\gg\mu_B$ (see above). Therefore,
at sufficiently large fields ($2\Delta_\orbit\gg\Delta_Z\gg
T_K^\mathrm{SU(4)}$), one has four split-Kondo peaks at
$\omega\approx\pm2\Delta_\orbit$ and $\omega\approx\pm\Delta_Z$;
see Fig.~\ref{su4cnt::fig:2} (left inset).  At moderate fields
such that
\begin{math}
2\Delta_\orbit\gtrsim T_K^\mathrm{SU(4)} \gg T_K^\mathrm{SU(2)}
\gtrsim \Delta_Z
\end{math},
one can have a three-peak structure; see Fig.~\ref{su4cnt::fig:2}
(dashed line). The lifted degeneracy in the orbital pseudo-spin
gives two side-peaks at $\omega\approx\pm2\Delta_\orbit$ while the
spin still retains a Kondo effect and gives the central peak.  The
central peak (which is now at $\omega=0$) corresponds to a
conventional SU(2) Kondo effect and hence is much narrower than
the central resonance for $B_{\parallel}=0$.

\begin{figure}
\centering%
\includegraphics*[width=85mm]{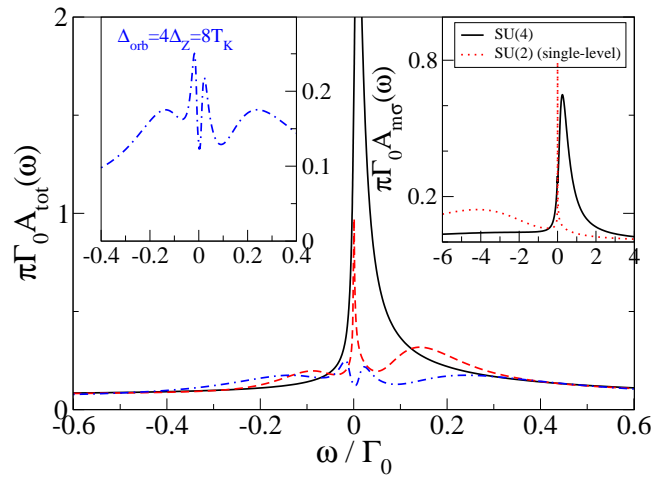}%
\caption{(Color online) \emph{NRG results for the total spectral density
    $A_\total(\omega)$}. The solid line is for the case of
  $\Delta_{\orbit}=\Delta_{Z}=0$, the dashed line for
  $\Delta_{\orbit}=8T_K^\mathrm{SU(4)}$
  ($T_K^\mathrm{SU(4)}=0.0133\Gamma_0$) with $\Delta_{Z}\approx 0$, and
  the dash-dotted line for
  $\Delta_\orbit=4\Delta_Z=8T_K^\mathrm{SU(4)}$.
  (Parameters: $\eps_d=-10\Gamma_0$, $U=200\Gamma_0$, $\Gamma_0=0.01D$,
  and $\delta\Gamma=0$). Left inset: Zoom of the four peak splitting in
  $A_\total(\omega)$.  Right inset: Comparison of the SU(4) (solid line)
  and the single-level SU(2) (dotted line) Kondo models. (Parameters:
  $\eps_d=-5\Gamma_0$, $U=500\Gamma_0$, $\Gamma_0=0.02D$, and
  $\delta\Gamma=0$).}
\label{su4cnt::fig:2}
\end{figure}

The above features of $A_\total(\omega)$ at equilibrium are directly
reflected in the non-linear conductance, $\mathcal{G}\equiv dI/dV$, an
experimentally measurable quantity. The current through the system $I$
can be expressed in terms of the local DOS~\cite{Meir-Wingreen}:
\begin{math}
I=(e/\hbar)\sum_{\sigma=\uparrow,\downarrow}\sum_{m=1,2}
\int{d\omega}\;[f_L(\omega)-f_R(\omega)][\Gamma_0(\omega)
+\delta\Gamma(\omega)]A_{m\sigma}(\omega)
\end{math},
where $\Gamma_0(\omega)=\pi\rho_0(\omega)|V_{0}|^2$,
$\delta\Gamma(\omega)=\pi\rho_0(\omega)|V_{X}|^2$, and $f_{L(R)}$
is the Fermi function for the left(right) lead.  The NRG procedure
is valid only at equilibrium so we need a method capable to
produce the non-equilibrium DOS and the non-linear current. We
choose to use a combination of the NCA and the EOM
methods~\cite{eom-nca}. Figures~\ref{su4cnt::fig:3}(a) and (b)
display the local DOS and the differential conductance,
respectively, for several $B_{\parallel }$. We take
$\Delta_\orbit$ ranging from $\Delta_\orbit=0.5T_K^\mathrm{SU(4)}$
to $\Delta_\orbit=1.5T_K^\mathrm{SU(4)}$ with
$\mu_{\orbit}=10\mu_B$ leading to $\Delta_Z=\Delta_\orbit/5$. For
the lowest magnetic field, $A_\total(\omega)$ exhibits two
splittings, namely the orbital and the Zeeman splittings with
peaks at $\omega\approx\pm2\Delta_{\orbit}$ and
$\omega\approx\pm\Delta_{Z}$, respectively, in excellent agreement
with the NRG calculations. As $\Delta_{\orbit}$ enhances beyond, a
substructure arises in the outer peaks. These new side-peaks at
$\omega \approx \pm 2\Delta_{\orbit}\pm\Delta_{Z}$ correspond to
the simultaneous \emph{spin-flip inter-orbital} transitions.
Increasing further $B_{\parallel}$ these side-peaks in the DOS get
better resolved~\cite{NRG}. All these features are also present in
the differential conductance plotted in
Fig.~\ref{su4cnt::fig:3}(b).

\begin{figure}
\centering %
%\rule{40mm}{40mm}
\includegraphics*[width=70mm]{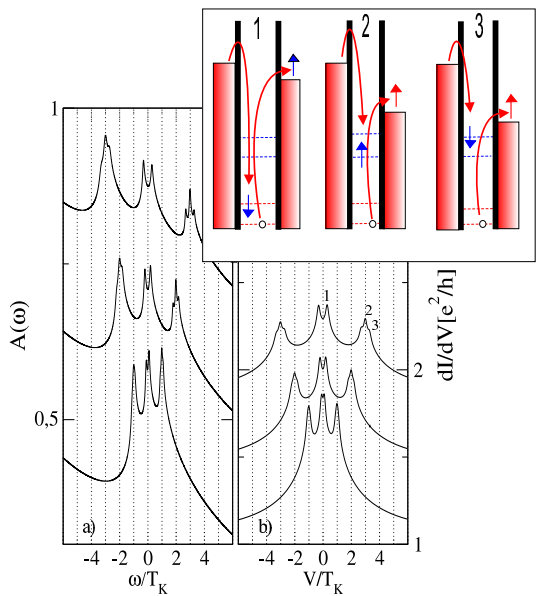}
\caption{(Color online). \emph{NCA+EOM results for the SU(4) Kondo
    model}. (a) $A_\total(\omega)$ and (b) $dI/dV$ versus $eV$ for
  different magnetic fields ranging from
  $\Delta_{\orbit}=0.5T_K^\mathrm{SU(4)}$
  (bottom) to $\Delta_{\orbit}=1.5T_K^\mathrm{SU(4)}$ (top). (The curves
  are shifted
  vertically for clarity). When $B_{\parallel}\neq 0$, the Kondo
  resonance splits due to the removal of both spin and orbital
  degeneracies.  The rest of parameters are: $\epsilon_d=-4\Gamma_0$ and
$T=0.003\Gamma_0$ with $\delta\Gamma=0$.  The quantum dot is
symmetrically coupled to two leads ($\Gamma_L=\Gamma_R=\Gamma_0$)
consisting of Lorentzian bands of width $2D=20\Gamma_0$. Inset:
schematic representation of the allowed transitions. 1)
intra-orbital \emph{with spin-flip}, 2) inter-orbital
\emph{without spin-flip} and 3) inter-orbital \emph{with
spin-flip}.} \label{su4cnt::fig:3}
\end{figure}

\begin{figure}
\centering %
\includegraphics*[width=85mm]{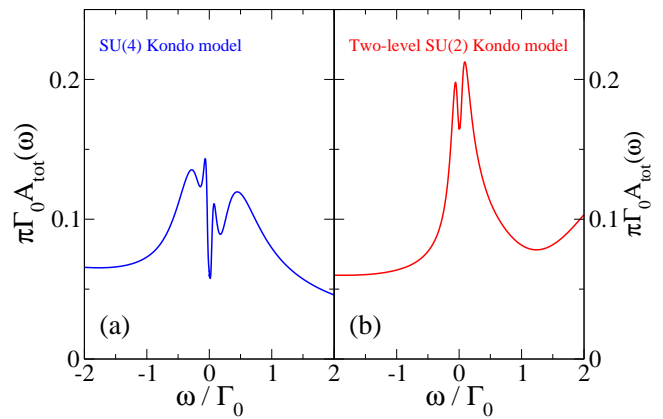}
\caption{(Color online). NRG results for the total spectral
density comparing (a) the SU(4) Kondo
  model ($\delta\Gamma=0$) and (b) the two-level SU(2) Kondo model
  ($\delta\Gamma=\Gamma_0$).  Parameters are
  $\Delta_\orbit=5\Delta_Z=15T_K^\mathrm{SU(4)}$ ($\approx
  0.2\Gamma_0$), $\eps_d=-10\Gamma_0$, $U=1000\Gamma_0$, and
  $\Gamma_0=0.01D$.}
\label{su4cnt::fig:4}
\end{figure}

\emph{Two-level SU(2) Kondo model at finite field}.---As already
shown, both the SU(4) and TL SU(2) Kondo models lead to similar
results for the Kondo temperature and the linear conductance.
Nevertheless, the inherent physics is completely different. In the
highly symmetric SU(4) Kondo model, the orbital pseudo-spin and
the real spin are indistinguishable (and screened simultaneously
at zero field). %%
On the contrary, in less symmetric multiple-level SU(2) Kondo
models, the spin should be clearly distinguished from the orbital
sector (tunneling processes preserve \emph{only the spin}). This
becomes clear at finite magnetic fields:
% Especially, the lift of
%the orbital degeneracy should be entirely different from that of
%spin degeneracy. %%
When $B_{\parallel}$ lifts the orbital degeneracy, only the lower
orbital level is occupied so the physics is essentially that of a
single level Kondo model~\cite{Boese02}. The Kondo resonance peak
gets narrower with increasing $B_{\parallel}$. Since
$B_{\parallel}$ also breaks the spin degeneracy, the resulting DOS
displays the usual Zeeman splitting. Overall, there are only two
peaks around $E_F$ (Fig.~\ref{su4cnt::fig:4}).
%All the features above are captured by our NRG
%calculations (see Fig.~\ref{su4cnt::fig:4} for a comparison).
%Therefore, one can conclude that the presence of $B_{\parallel}$
%provides a fundamental test for the occurrence of SU(4) Kondo
%physics in carbon nanotubes.

\emph{Summary}.--- We have demonstrated that quantum fluctuations
between the orbital and the spin degrees of freedom in carbon
nanotube QD´s result in an SU(4) Kondo effect at low temperatures.
This exotic Kondo effect manifests as a four-peak splitting in the
non-linear conductance when an axial magnetic field is applied.
Similar effects may appear in other systems like vertical
dots~\cite{Sasaki04a} where the orbital quantum number is
preserved during tunneling. Recent transport experiments in
CNT´s~\cite{Jarillo-Herrero04z} clearly support our theoretical
findings.
%We also mention recent experiments with
%vertical dots, another system where the orbital quantum number may
%be preserved during tunneling, where enhanced Kondo effects have
%been reported due to the additional orbital
%degeneracy~\cite{Sasaki04a}.

%% On the other hand, the $dI/dV$ data in the presence of $B_{\parallel}$
%% in Ref.~\cite{Jarillo-Herrero04z} strongly suggests that the system
%% bears the SU(4)-symmetric Kondo nature.

%% Recently, it has been reported that the existence of both spin and
%% orbital degeneracy in vertical dots gives rise to a enhanced Kondo
%% effect~\cite{Sasaki04a}.  On the ground discussed above, however, the
%% results in Ref.~\cite{Sasaki04a} does not rule out the TL SU(2)
%% Kondo effect as their underlying physics. In principle, such a strongly
%% correlated state could be ascribed either to the TL SU(2) Kondo
%% effect or to the SU(4) Kondo state.  On the other hand, the $dI/dV$ data
%% in the presence of $B_{\parallel}$ in Ref.~\cite{Jarillo-Herrero04z}
%% strongly suggests that the system bears the SU(4)-symmetric Kondo
%% nature.

We thank Pablo Jarillo-Herrero, Silvano de Franceschi and Leo
Kouwenhoven for sharing their experimental results with us prior
to publication and for many helpful discussions. We also thank
Markus B\"uttiker, Karyn Le Hur and David S\'anchez for fruitful
discussions. Work supported by the EU RTN No.HPRN-CT-2000-00144,
the Spanish MECD and MCyT through grant MAT2002-02465 and the
"Ram\'on y Cajal" program, the SKORE-A, and the eSSC at Postech.

%%%%%% References
%% \bibliographystyle{physrev}%
%% \bibliography{aliases,cond-mat,staphy,quaphy,physics,science,mathematics,%

\begin{thebibliography}{10}
\bibitem{NRG}
The NRG calculations in equilibrium also show a tiny substructure
around the $\omega \approx \pm 2\Delta_{\orbit}\pm\Delta_{Z}$
indicating the occurrence of spin-flip inter-orbital transitions.
However, due to the broadening of the outer peaks we are not able to
resolve the peaks inside.
\end{thebibliography}
%%   su4cnt,opus}%

\end{document}